\newcommand{\Lsun} {L$_\odot$}

\documentclass[11pt]{article}
\usepackage{moriond,epsfig}

\def\be{\begin{equation}}
\def\ee{\end{equation}}
\def\bea{\begin{eqnarray}}
\def\eea{\end{eqnarray}}

\begin{document}
\vspace*{4cm}
\title{DUST EMISSION BY AGN AND STARBURSTS}

\author{ Ralf Siebenmorgen (rsiebenm@eso.org)}

\address{European Southern Observatory, Karl-Schwarzschildstr. 2, \\
        D-85748 Garching b. M\"unchen}

\maketitle\abstracts{
Present AGN and starburst models aiming to account for the observed
infrared SEDs consider a physical description of the dust and a
solution of the radiative transfer problem.  Mid infrared spectra
obtained at different spatial scales (SST-IRS, ISO and Timmi2) are
presented. They show that PAH bands are detected in starburst regions
but significantly reduced near the centre of AGN.  This is explained
by examining the heating mechanism of PAHs after hard (FUV, X-ray)
photon interactions.  Most economic radiative transfer models of
starbursts and AGN are presented where only key parameters such as
luminosity, dust mass or size of the nucleus are varied. For both
activity types syntetic spectra are made available.  The successful
application of the starburst model is demonstrated by fitting broad
band data and detailed Spitzer IRS spectra of NGC7714. The AGN model
is applied to ISOCAM and ISOPHT broad band data of a sample of 68
radio galaxies and quasars of the 3CR catalogue. Radiative transfer
models of galaxies with Hidden Broad Line Regions (HBLR) are
presented. Their SED enable us to separate the contributions from the
dusty disc of the AGN and the starbursts.  We find that the
combination of AGN heated discs and starbursts provide good fits to
the data.  The composite model is consistent with the unified scheme
and the idea that the infrared emission of AGN is dominated by a dusty
disc in the mid-infrared and starbursts in the far-infrared.
According to the unified scheme, AGN are surrounded by a dust-torus,
and the observed diversity of AGN properties results from the
different orientations relative to our line of sight.  The strong
resonance of silicate dust at 10 $\mu$m is therefore, as expected,
seen in absorption towards many type-2 AGN.  In type-1 AGN, it should
be seen in emission because the hot inner surface of the dust torus
becomes visible. However, this has not been observed so far, thus
challenging the unification scheme or leading to exotic modifications
of the dust-torus model.  Here the recent discovery of the 10~$\mu$m
silicate feature in emission in luminous quasar spectra observed with
the SST/IRS is reported. }

\section{Introduction}

Nuclei of luminous infrared galaxies are generally dust enshrouded and
not transparent.  Therefore radiative transfer calculations have to be
carried out for an optically thick dusty medium. This has been done in
various approximations of the physics in starburst and AGN dominated
galaxies. In the following I summarize some of our recent findings on
the dust interaction with hard radiation, the application of the
models to recent Spitzer data and report on the discovery of the
silicate emission in type 1 AGN.

\begin{figure} [htb]
\vspace{-0.cm}
\centerline {\psfig{figure=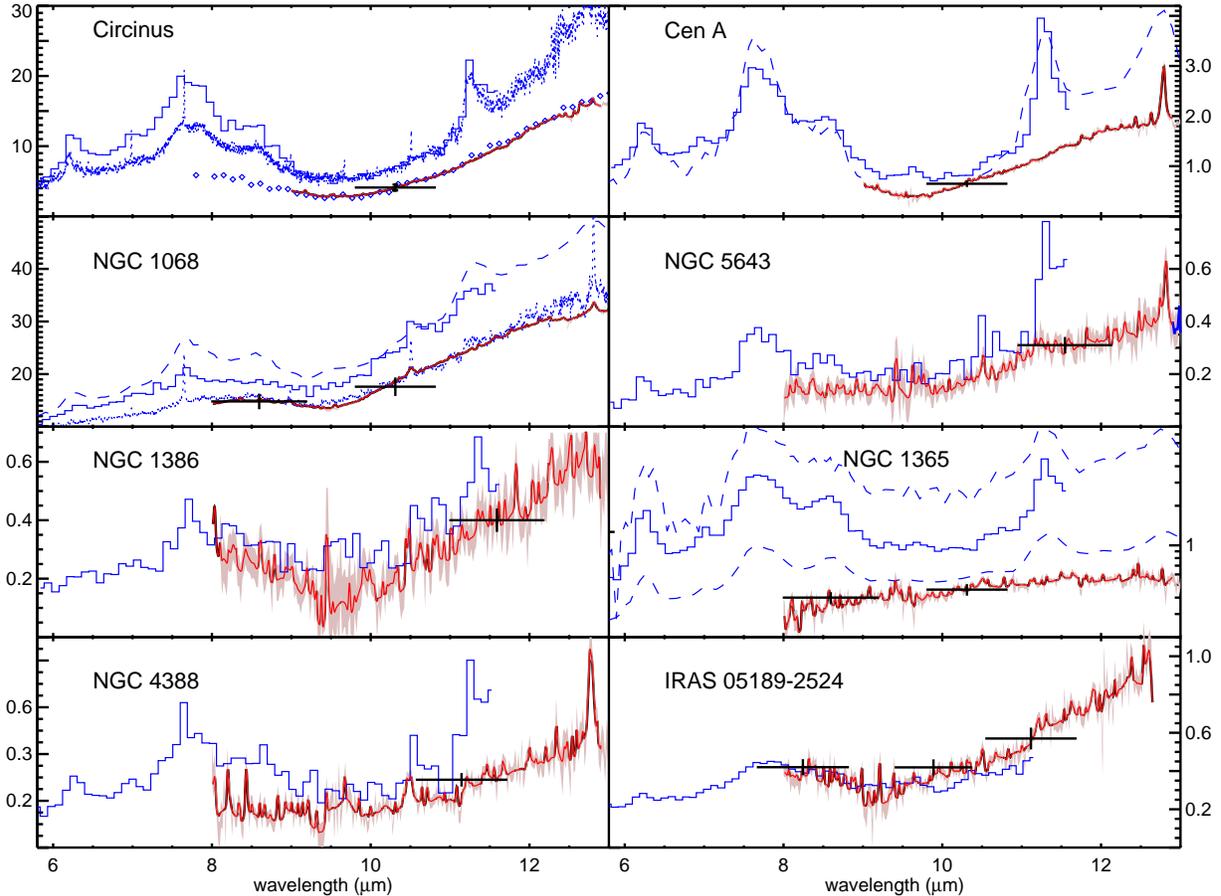,width=14.cm,angle=90}}
  \caption{Mid-infrared flux densities in Jy of Seyfert galaxies
  (Siebenmorgen et al. 2004 [2]): histograms and dashed lines are
  large beam (14$''-20''$) ISO observations, thick lines are narrow
  slit (3$''$) data with TIMMI2. Note that PAH bands are visible in
  the large beam spectra but absent in the high spatial resolution
  data of the same galaxy. \label{fig.1}}
\end{figure}

\section{The Dust Model}

Any dust model should be based on fitting interstellar extinction
curves, abundance constrains and dust features. We use large carbon
and silicate grains with radii, $a$, between 300 and 2400\AA \/ with a
size distribution $n(a) \propto a^{-3.5}$; very small graphites ($a =
10$\AA); and two kinds of PAHs (30 C and 20 H atoms; 252 C and 48 H
atoms). The PAH abundance is such that for 10$^5$ protons in the gas
phase one finds one carbon atom in each PAH population. The infrared
absorption coefficient, $\kappa_{\lambda}$, of astronomical PAH is
derived by Siebenmorgen et al. (2001) ~[1].

The MIR emission bands give important insight on the activity type of
a galaxy: starburst or AGN. Active galaxies with Seyfert nuclei often
show strong PAH emission bands when observed at large scales, say with
the spatial resolution of ISO/SST ($\ge 10''$). However, mid infrared
spectra of the same galaxy are PAH free when observed at high spatial
resolution and where the slit is centered close to the nucleus. A few
examples of this finding are presented in Fig.\ref{fig.1}.

This observational fact and a study of the survival of PAHs in
starburst and AGN environments with energetic photons of, at least, a
few up to hundreds of eV is presented by Siebenmorgen et al. (2004)
[2]. It is found that small grains evaporate after interaction with
hard photons; in particular with photons of more than 50eV, which can
only be emitted by X--ray sources. A simple explanation is given by
studying the quantum statistical behaviour of PAHs. This destruction
process is considered in the radiative transfer models described
below.

% -------------------------------

\section{Radiative transfer models}

The radiative transfer in a dust cloud around a starburst is computed
following Siebenmorgen et al. (2001)~[1], around an AGN after
Siebenmorgen et al. (2004b) [3].  The major difference between both
radiative transfer models is that AGN are heated by a central source
emitting hard photons whereas starburst galaxies are heated by stars
which are distributed over a large volume.

\begin{figure} [b]
\centerline {\psfig{figure=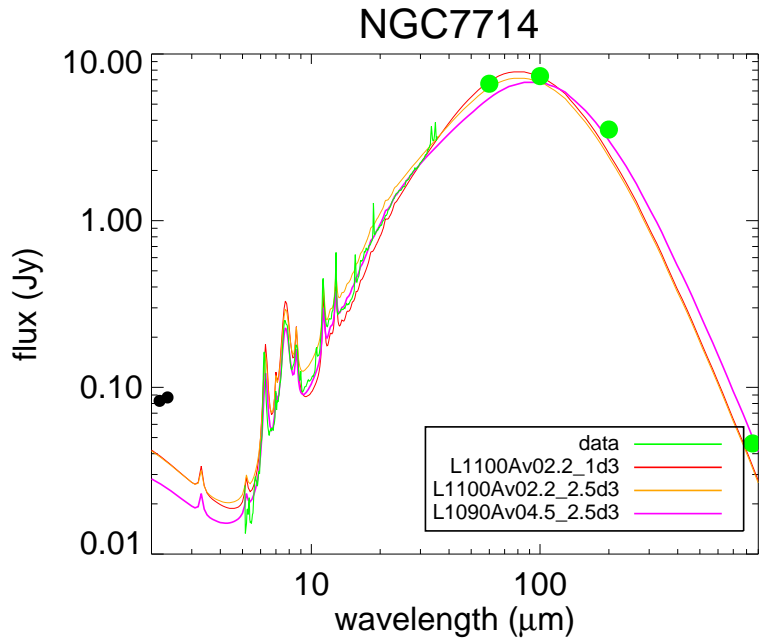,width=11cm}}
\caption{Spectral energy distribution of the starburst galaxy NGC7714. 
Spitzer IRS (Charmandaris, priv. com.) and NED data are in green,
other curves are starburst models with luminosity between $10^{10.9
... 11}$\Lsun (by assuming a distance of 40Mpc), visual extinction $A_V
= 2 ... 5$mag and a dust density surrounding the OB stars of $n({\rm
H})$: $10^3$, $2.5 \times 10^3$\,cm$^{-3}$. The size of the nucleus is
not varied and fixed to 3kpc. \label{fig.2}}
\end{figure}

The radiative transfer in starburst nuclei is solved by a scheme first
described by Kr\"ugel \& Tutokov (1978) [4] for the Galactic Centre
and generalized by Kr\"ugel \& Siebenmorgen (1994) [5] for
starbursts. In the starburst models two populations of stars are
considered OB stars have a uniform luminosity of $20\,000$\Lsun\ and a
stellar temperature $T_* = 25\,000$\,K and KM stars with stellar
temperature $T_* = 4\,000$\,K. The stellar density of OB stars changes
with galactic radius $r$ from the center like $\propto 1/r$.  An
important feature of the starburst model is that each OB star is
surrounded by a dust shell, for which we apply a constant dust
density, $n^{OB}({\rm H})$.  The inner radius of this circumstellar
dust shell is given by the photo--destruction or evaporation radius of
the grains.  The outer radius of such {\it {hot spots}} is determined
by the condition of equal heating of the dust from the star and from
the radiation field in the galactic nucleus.  Therefore most of the UV
light of a particular OB star is absorbed by the circumstellar dust
shell and then re-emitted at infrared wavelength into the galactic
nucleus. Hence for starburst galaxies the radiative transfer has to be
solved for two different scales: first, on galactic scale of the
nucleus and second, for each of the dust embedded OB stars. Both
transfer problems are linked to each other via the boundary conditions
(for example the outer radius of the circumstellar shell) and an
iterative scheme is applied to find a self consistent
solution. Starburst models presented by other groups often simplify
matters and consider the first problem for an ensemble of OB stars or
molecular clouds assuming an optically thin medium surrounding it.  A
grid of our starburst model spectra is available where four basic
parameters are varied: the luminosity of the OB and KM stars, the size
of the nucleus, the total extinction of the nucleus (as measured from
surface to center), and the dust density of the circumstellar shells
surrounding the OB stars. In Fig.~\ref{fig.2} we present one example
of such model fits to the SED of NGC7714. The fits successfully
reproduce the detailed Spitzer IRS spectrum (Charmandaris, priv. com.)
of the dust emission. Gas features are not treated. Slight fine tuning
of the PAH abundance will result in an even better fit.

% -------------------------------

% -------------------------------

\begin{figure} [tb]
\centerline {\psfig{figure=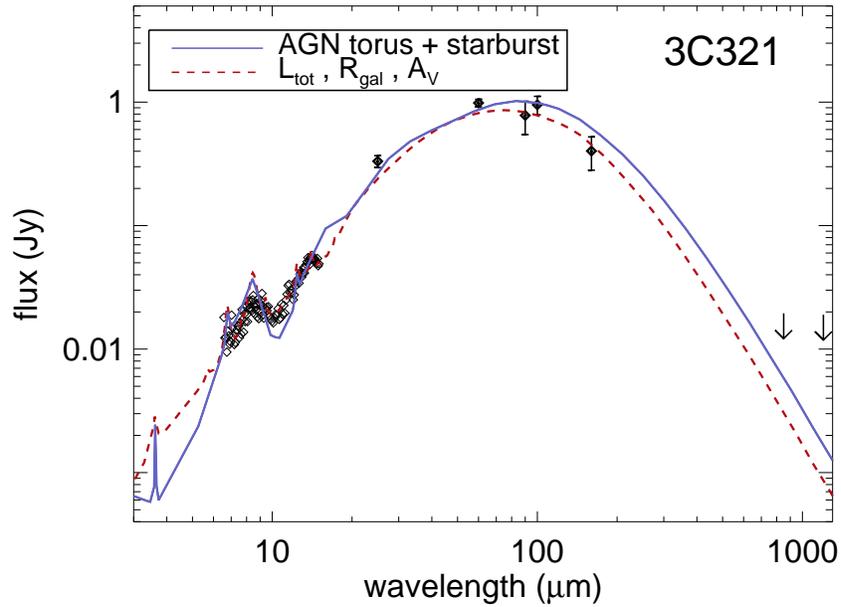,width=9cm,angle=90}}
\caption{Spectral energy distribution of the hidden broad line region 
galaxy 3C321. Two radiative transfer models are compared: i) a three
parameter fit (dashed line) using the AGN model by Siebenmorgen et
al. (2004b) [3] and ii) a combination of a tapered torus model with
starburst activity by Efstathiou \& Siebenmorgen (2005) [6].  Both
models are consistent with the observations. However, the later is in
better agreement with the unification model and provides correction
for the anisotropic torus emission. \label{fig.3}}
\end{figure}

% -----------------------------------------------------------------------
\begin{figure} [htb]
\hbox{\hspace{0cm}
\psfig{file=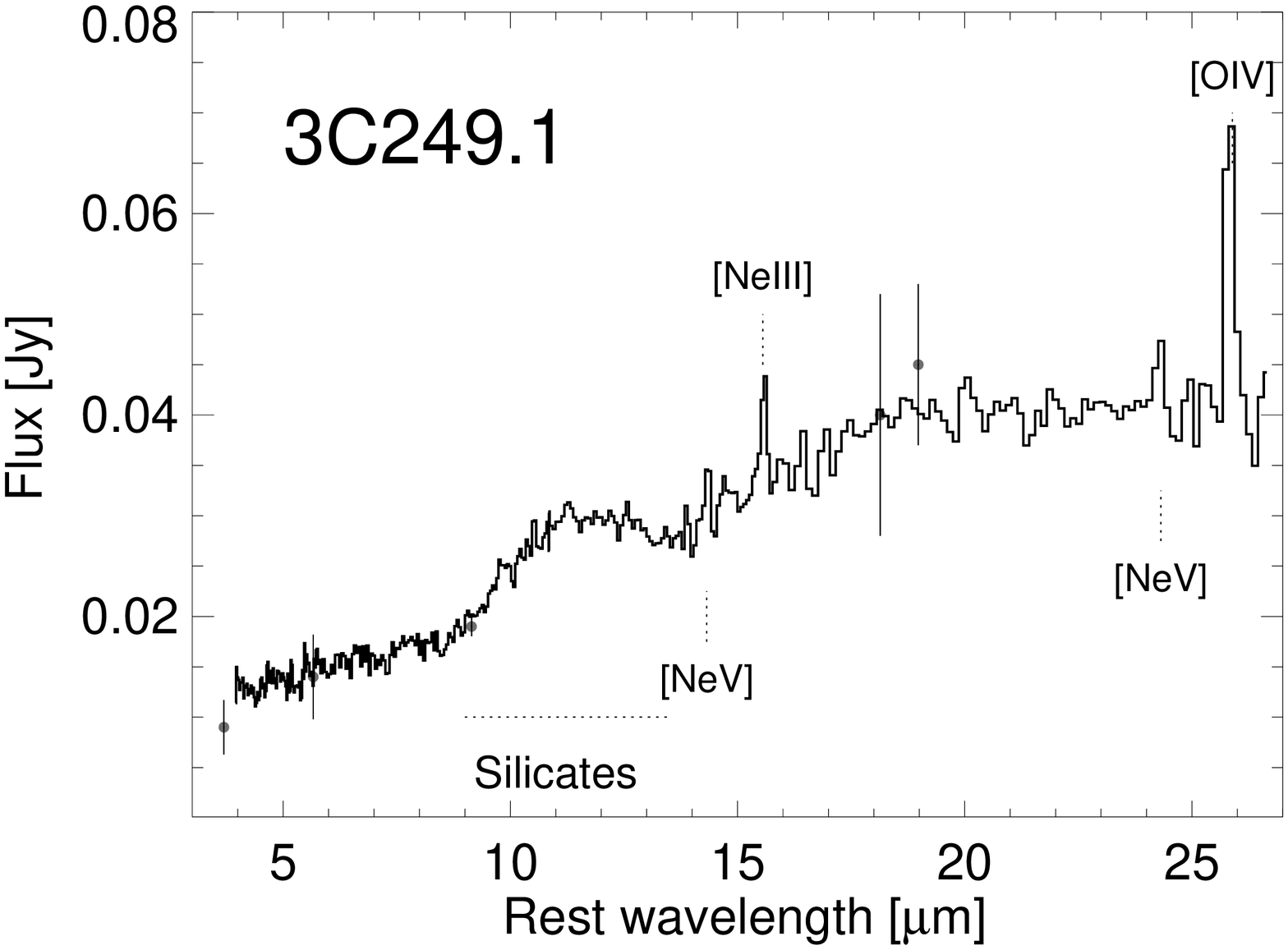,angle=90,width=8.cm,height=8cm}
\hspace{0.cm}
\psfig{file=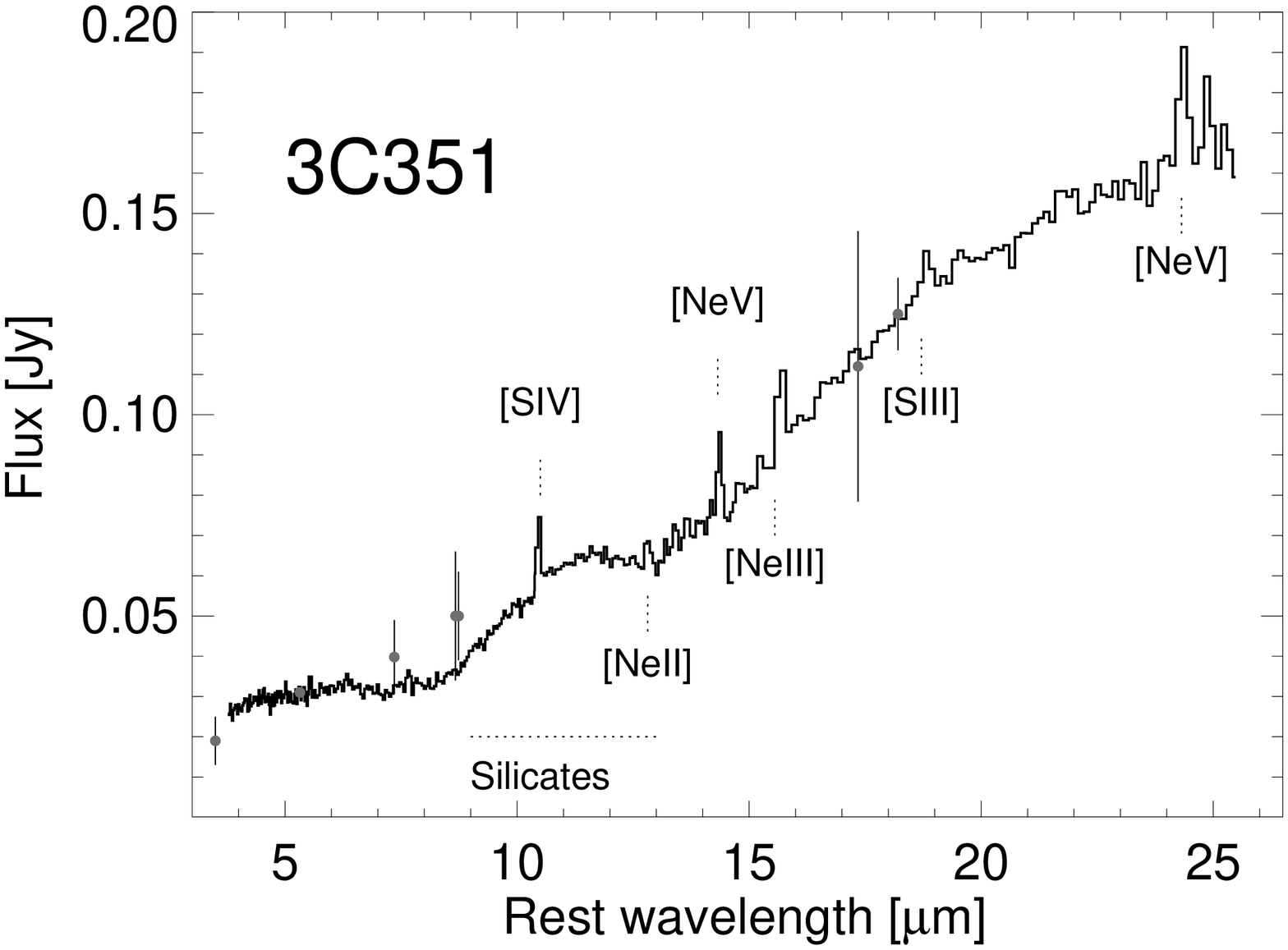,angle=90,width=8.cm,height=8cm}}
\vspace{0cm}
\hbox{\hspace{0cm}
\psfig{file=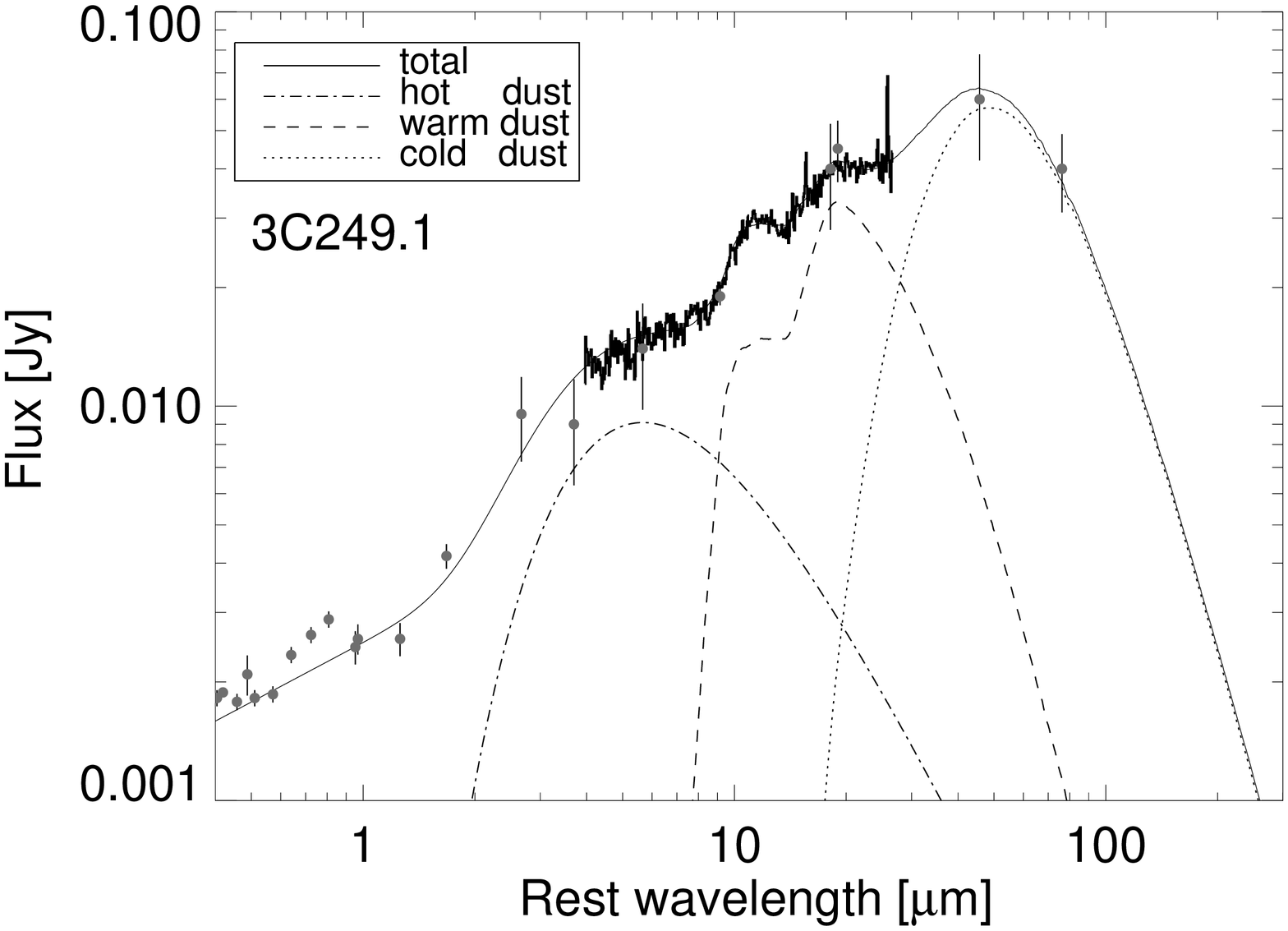,angle=90,width=8.cm,height=7cm}
\hspace{0.cm}
\psfig{file=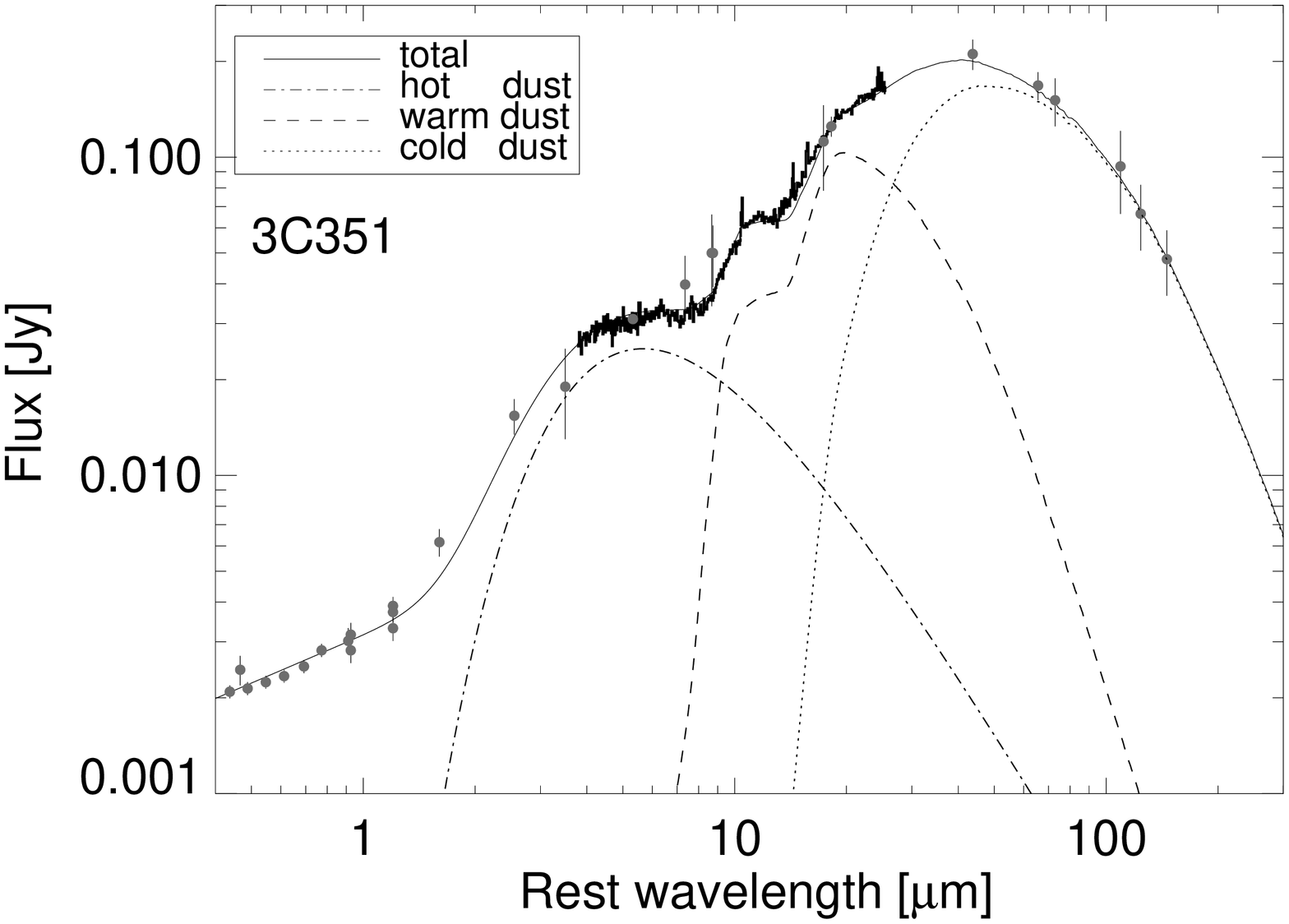,angle=90,width=8.cm,height=7cm}}
\caption{Spectra and models of the quasars 3C249.1 and 3C351, obtained with the
Infrared Spectrograph (IRS) of the Spitzer Space Telescope
(Siebenmorgen et al. (2005) [7]). AGN-typical high-excitation emission
lines like [\,Ne~~V\,] $\lambda$=14.3$\mu$m,$\lambda$=24.3$\mu$m are
marked with vertical dotted lines. The broad bump in the wavelength
range 9 -- 13 $\mu$m, indicated by horizontal dotted lines, is the
detected silicate emission.  The models consist of a central heating
source with an X-ray-to-infrared power-law spectrum, two dust
components and a black body. The warm (140 -- 200K) dust component is
located 30 -- 100pc from the central heating source, while the cold
(30 --70K) dust is more than 600pc and up to a few kpc further
out.\label {fig.si}}
\end{figure}

\clearpage

\section{AGN and starburst composite spectra}

Most active galaxies are known to be composite objects where a Seyfert
nucleus is surrounded by a ring of star formation.  Efstathiou \&
Siebenmorgen (2005) [6] fit ISO data of galaxies with hidden broad
line regions.  In their models a combination of a tapered AGN torus
with starburst activity is treated using six free parameters.  One
example of the model fit is given for 3C321 in Fig.\ref{fig.3}. We
compare the fit with a pure and simplified AGN model by Siebenmorgen
et al. (2004b) [3] where only 3 parameters: luminosity, effective size
and extinction of the nucleus are varied to compute the SED.  (A full
grid of such models is available by request.)  As shown in
Fig.~\ref{fig.3} , both models are consistent with the data. However,
the model by Efstathiou \& Siebenmorgen enables to separate the
contribution from the dusty disc of the AGN and the dusty starbursts.
We find that tapered discs dominate the emission in the mid infrared
part of the spectrum and the starbursts in the far infrared. The AGN
models by Efstathiou \& Siebenmorgen have a two dimensional structure
therefore they provide a correction factor of the AGN luminosity for
anisotropic emission and are more consistent with the unified scheme.

% ----------------------------------

\section{AGN unification and the MIR emission of quasars}

According to the unified scheme, AGN are surrounded by a dust-torus,
and the observed diversity of AGN properties results from the
different orientations relative to our line of sight.  The strong
resonance of silicate dust at 10$\mu$m is therefore, as expected, seen
in absorption towards many type-2 AGN.  In type-1 AGN, it should be
seen in emission because the hot inner surface of the dust torus
becomes visible. However, this has not been observed so far, thus
challenging the unification scheme or leading to exotic modifications
of the dust-torus model.  In Fig.\ref{fig.si} we present our recent
discovery of the 10~$\mu$m silicate feature in emission in luminous
quasar spectra observed with the Infrared Spectrograph of the Spitzer
Space Telescope is presented (Siebenmorgen et al. (2005) [7]).

The Spitzer observations, as well as the photometric data at other
infrared wavelengths, can be reproduced by a model consisting of three
components: cold (30 -- 70K) and warm (140 -- 200K) dust and a hot
blackbody. The primary heating source has a power-law spectrum,
F$_{\nu}$\,$\propto$\,$\nu$$^{\rm -0.7}$, in the wavelength range 1\AA
\/ -- 15~$\mu$m.  The emission of the dust is treated to be optically
thin which is a reasonable assumption for a face-on viewed quasar.  Of
course, this approach needs further refinement by a self-consistent
axial-symmetric radiative transfer model of the quasar emission, on
which we are working.

% -----------------------------------------------------------------------

\end{document}